%% file: nisq-enum.tex
\documentclass[a4paper,accepted=2023-02-12]{quantumarticle}
\pdfoutput=1

\def\issubmission{0}
\def\isfullversion{0}
\def\buildexternal{1}

\usepackage[l2tabu,orthodox]{nag}

\usepackage{subcaption}
\usepackage{microtype}
\usepackage[bookmarksdepth=2]{hyperref}
\usepackage{booktabs}  
\usepackage{comment}
\usepackage{myCustomMathsSymbols}
\usepackage{float}
\usepackage[inline]{enumitem}
\setlist[enumerate]{label=(\roman*)}
\usepackage{soul}
\usepackage[dvipsnames]{xcolor}
\usepackage[numbers]{natbib}

\usepackage{xcolor}
\definecolor{oxygenorange}{HTML}{FFDD00}
\usepackage[color=oxygenorange]{todonotes}

\usepackage{ifthen}
\newcommand{\submission}[2]{%
\ifthenelse{\equal{\issubmission}{1}}%
{#1}%
{#2}%
}%

\newcommand{\fullversion}[2]{%
\ifthenelse{\equal{\isfullversion}{1}}%
{#1}%
{#2}%
}%

\usepackage{algorithm2e}
\usepackage{amsmath,amsfonts,amssymb,amsthm}
\newtheorem{lemma}{Lemma}
\newtheorem{theorem}{Theorem}
\newtheorem{proposition}{Proposition}
\newtheorem{corollary}{Corollary}

\usepackage{xspace}
\usepackage[lambda,landau,operators,probability,sets,logic,complexity,asymptotics]{cryptocode}
\usepackage{dsfont}

\renewcommand{\vec}[1]{\ensuremath{\boldsymbol{#1}}\xspace}
  \newcommand{\mat}[1]{\ensuremath{\boldsymbol{#1}}\xspace}

\newcommand{\radenum}{A}

\DeclareMathOperator{\gh}{gh}
\DeclareMathOperator{\vol}{vol}
\DeclareMathOperator{\Span}{span}

\DeclareMathOperator{\mR}{\mathbb{R}}
\DeclareMathOperator{\mZ}{\mathbb{Z}}
\DeclareMathOperator{\cL}{\mathcal{L}}

\newcommand{\SVP}{\textsf{SVP}}

\usepackage{tikz,pgfplots}
\usetikzlibrary{pgfplots.groupplots}
\pgfplotsset{compat=newest}
\usetikzlibrary{calc}
\usetikzlibrary{arrows}

\usetikzlibrary{external}
\tikzset{external/optimize=false}
\tikzset{external/export=false}

\newcommand{\tikzexternalizemaybe}{%
  \ifthenelse{\equal{\buildexternal}{1}}%
  {\tikzset{external/export=true}}%
  {\tikzset{external/export=false}%
}}%

\tikzset{yixin lattice point/.style={draw,circle,color=gray,fill,inner sep=#1}}
\tikzset{yixin lattice point/.default=2pt}

\definecolor{DarkPurple}{HTML}{332288}
\definecolor{DarkBlue}{HTML}{6699CC}
\definecolor{LightBlue}{HTML}{88CCEE}
\definecolor{DarkGreen}{HTML}{117733}
\definecolor{DarkRed}{HTML}{661100}
\definecolor{LightRed}{HTML}{CC6677}
\definecolor{LightPink}{HTML}{AA4466}
\definecolor{DarkPink}{HTML}{882255}
\definecolor{LightPurple}{HTML}{AA4499}
\definecolor{DarkBrown}{HTML}{604c38}
\definecolor{DarkTeal}{HTML}{23373b}
\definecolor{LightBrown}{HTML}{EB811B}
\definecolor{LightGreen}{HTML}{14B03D}

\pgfplotsset{width=1.0\textwidth,
  height=0.45\textwidth,
  cycle list={%
    solid,\\%
    dotted,DarkBlue,very thick\\%
    dashed,DarkGreen,thick\\%
    loosely dotted,very thick,DarkRed\\%
    loosely dashed,black,very thick,DarkBrown\\%
    loosely dashdotted,darkgray,very thick,DarkTeal\\%
    \\%
  },
  legend pos=north west,
  legend style={fill=none},
  legend cell align={left}}

\usepackage{attachfile}

\usepackage{pgfkeys}
\input{constants}

\usepackage{listings}
\lstdefinelanguage{Sage}[]{Python}{morekeywords={True,False,sage,cdef,cpdef,ctypedef,self},sensitive=true}
\usepackage{navigator}
        
\hypersetup{colorlinks=true,citecolor=darkgray,linkcolor=darkgray}

\usepackage[capitalize,noabbrev]{cleveref}

\title{Variational quantum solutions to the Shortest Vector Problem}
\author{Martin R.~Albrecht}
\affiliation{King's College London and SandboxAQ. email: \texttt{martin.albrecht@kcl.ac.uk}}
\author{Milo\v{s} Prokop}
\affiliation{University of Edinburgh. email: \texttt{m.prokop@sms.ed.ac.uk}}
\author{Yixin Shen}
\affiliation{Royal Holloway, University of London. email: \texttt{yixin.shen@rhul.ac.uk}}
\author{Petros Wallden}
\affiliation{University of Edinburgh. email: \texttt{petros.wallden@ed.ac.uk}}

\begin{document}
\setlength{\intextsep}{5pt plus 2pt}
\maketitle

\begin{abstract}
  A fundamental computational problem is to find a shortest non-zero vector in Euclidean lattices,  a problem known as the Shortest Vector Problem (SVP).  This problem is believed to be hard even on quantum computers and thus plays a pivotal role in post-quantum cryptography. In this work we explore how (efficiently) Noisy Intermediate Scale Quantum (NISQ) devices may be used to solve SVP\@.
  Specifically, we map the problem to that of finding the ground state of a suitable Hamiltonian. In particular,
  \begin{enumerate*}
  \item we establish new bounds for lattice enumeration,  this allows us to obtain new bounds (resp.~estimates) for the number of qubits required per dimension for any lattices (resp.~random \(q\)-ary lattices) to solve SVP;
  \item we exclude the zero vector from the optimization space by proposing (a)  a different classical optimisation loop or alternatively (b) a new mapping to the Hamiltonian.
  \end{enumerate*}
  These improvements allow us to solve SVP in dimension up to 28 in a quantum emulation, significantly more than what was previously achieved, even for special cases. Finally, we extrapolate the size of NISQ devices that is required to be able to solve instances of lattices that are hard even for the best classical algorithms and find that with $\mathbf{\approx 10^3}$ qubits such instances can be tackled.
\end{abstract}

\section{Introduction}\label{sec:introduction}
\input{introduction}
\section{Preliminaries}\label{sec:preliminaries}
\input{preliminaries}

\subsection{Related Work}\label{sec:relatedwork}
\input{relatedWork}

\section{Direct approaches to SVP with Variational Quantum Algorithms}\label{sec:svpWithVQA}
\input{mappingSVPtoHamiltonian}

\section{Bounds on lattice enumeration and application to variational quantum algorithms for solving SVP }\label{sec:compositeSVP}
\input{compositeSVP}





\subsection{General bounds on lattice enumeration}\label{sec:optBound}
\input{optimalBounds}

\section{Experimental results}\label{sec:expResults}
\input{experimental_results}


\submission{}{
\section*{Acknowledgements}
\input{acknowledgements}
}

%
%
%
%
%
%
%



\clearpage
\bibliographystyle{quantum}
\bibliography{abbrev3,crypto_crossref,local}

\appendix

\section{Techniques to optimise towards first excited state of problem Hamiltonian}\label{app:avoidZero}
\input{appendix_zeroVect}

\section{Probability of including the shortest lattice vector in the search space using naive qubit mapping strategies}\label{app:groundStateProbs}
\input{appendix_naive_qubit_assignment.tex}

\section{Conditional value at risk experiment data}\label{app:conditionalValAtRisk}

\input{appendix_cvar}
\input{appendix_proofs}

\clearpage
\onecolumn
\input{experimental_results/probGroundState.tex}

\input{appendix_cvar_fig}

\end{document}

%% file: constants.tex
\pgfkeys{
  /foo/bar/.initial=1,
}


%% file: introduction.tex
Cryptography studies the limits of computing: what can and cannot efficiently be computed. In 1976 Diffie and Hellman significantly expanded the realm of the possible by inventing public key cryptography~\cite{dif77} which allows two parties to agree on a shared secret over a public channel in the presence of a wiretapping adversary.\footnote{If the public channel is authenticated, the adversary may even actively interact with both parties.} Since its invention public-key cryptography has seen widespread adoption and is now a crucial building block for securing, say, the Internet. However, this advance was not unconditional but relies on the presumed hardness of some computational problem. Virtually all currently deployed public-key encryption schemes rely on the difficulty of one of two computational problems: the discrete logarithm problem and the problem of factoring large integers.

Everything changed in 1994 when Peter Shor's seminal work~\cite{shor97} showed that quantum computers could effectively solve those two central problems. While the timeline for when sufficiently big quantum computers may be available is uncertain, the proposed such timelines and the threat of ``collect-now-decrypt-later''-style attacks provoked global efforts to develop ``post-quantum cryptography'', cryptographic schemes that run on classical computers but resist attacks with quantum computers. The centre of these international efforts is the Post-Quantum Standardisation Process (NIST PQC) by the US National Institute of Standards and Technology (NIST)~\cite{NIST}.
To date, there are several candidates for post-quantum cryptography, mainly lattice-based, code-based, hash-based, multivariate cryptography and supersingular elliptic curve isogeny cryptography. Lattice-based cryptography seems to be a prime contender for large scale adaption: among the winners the NIST PQC process, three out of four are based on lattices.

While, of course, all post-quantum candidates are conjectured to be hard also on a quantum computer, a pressing question for adoption is ``how hard?''. That is, in order to pick parameters that are both efficient in practice but quantifiably resist attacks, the research community studies the quantum resources required to solve the underlying hard problems.

On the other hand, it is expected that the transition to quantum computers will start with a phase referred to as \emph{Noisy Intermediate Scale Quantum} (NISQ), featuring devices that consist of at most one thousand erroneous qubits. This low number of qubits makes use of any known error-correction technique infeasible, something that also puts stringent restrictions on the depth a quantum computation can have before the noise becomes dominant leading to essentially random outcomes. To overcome this limitation of these devices, hybrid classical-quantum algorithms are designed specifically for these devices while most of the famous quantum algorithms like Grover's search, Quantum Fourier Transform or Shor's algorithms are impracticable. A natural question is thus ``how hard are lattice problems on NISQ devices?''\footnote{By NISQ devices, in this paper, we mean devices with limited number of qubits and using algorithms with limited depth of quantum computation in each iteration. Of course, the limitation in the depth comes from the existence of ``noise'' but the effects of noise extend further, e.g. by reducing the probability that the classical optimisation loop ends-up with the correct solution. Since we mainly focus on the qubit-count and do not address the scaling of the probability of success even in the noiseless case, we also leave the full analysis of the effects of noise to a future publication.}. This question sheds light on the performance of these devices on a natural and central computational problem.

\paragraph{Contributions.}
After some preliminaries (\cref{sec:preliminaries}), we determine a suitable mapping of the central hard problem on lattices, the \emph{Shortest Vector Problem} (SVP) into the ground state of a Hamiltonian operator, the form required by the classical-quantum optimisation algorithms of our interest. In \cref{sec:svpWithVQA}, we introduce Ising spin Hamiltonian operators and explain how to map SVP to an optimisation problem in this framework. We discuss the challenges that arise with this formulation and give a solution to the ``zero vector problem'' for VQE\@.\footnote{In \cref{app:avoidZero} we outline a less efficient solution to impose the constraint at the Hamiltonian level adding new variables. That approach is applicable to QAOA, quantum annealing and adiabatic quantum computing thus extending our method.} Our problem formulation requires to know a priori bounds for each coordinate of a shortest vector in a given basis $\mat{B}$ of the lattice. In \cref{sec:compositeSVP} we thus analyse the cost of solving SVP in a non-adaptive exhaustive search which allows us to quantify the search space. In particular, we show that the size of the search space depends on the norms of the vectors forming the ``dual basis'' $\widehat{\mat{B}}\coloneqq {(\mat{B} \cdot \mat{B}^{T})}^{-1}\cdot \mat{B}$. This allows us to obtain a NISQ quantum algorithm to compute an HKZ reduced basis, one of the strongest notion of reduction, using $\tfrac{3}{2}n\log_2n+O(n)$ qubits and thus solving SVP in passing. This cost of non-adaptive enumeration was previously known only for a special class of lattices. We also perform extensive classical numerical experiments to study the average case behaviour of lattice reduction in the context of our NISQ quantum enumeration algorithm. In \cref{sec:expResults} we then show that our bounds allow us to run quantum emulation experiments that using up to 28 qubits are able to solve SVP in dimension 28 which is considerably more than prior literature. Extrapolating our experimental data we find that between \(1,000\) and \(1,600\) qubits suffice to encode SVP for a dimension 180 lattice, the current record dimension for ``Darmstadt SVP Challenge''~\cite{EC:DucSteWoe21}. For the avoidance of doubt, our results do not violate any previous claims on the hardness of lattice problems on quantum computers because in general we may hope for a running time at best \(2^{\lambda/2 + o(\lambda)}\) for instances encoded in \(\lambda = \tfrac{3}{2}n\log_2n+O(n)\) qubits.\footnote{Classically, SVP can be solved in time \(n^{1/(2e)n + o(n)}\) and \(\poly[n]\) memory or \(2^{0.292\,n + o(n)}\) time and memory.}

%% file: preliminaries.tex
\paragraph{Lattices.} A (Euclidean) \emph{lattice} $\cL$ is a discrete subgroup of $\mR^{d}$,
or equivalently the set $\cL(\vec{b}_{1},\dots,\vec{b}_{n})=\left\{ \sum_{i=1}^{n}x_{i}\cdot \vec{b}_{i} ~:~ x_{i}\in\mZ\right\} $
of all integer combinations of $n$ linearly independent vectors
$\vec{b}_{1},\dots,\vec{b}_{n} \in \mR^{d}$. 
Such $\vec{b}_i$ form a \emph{basis} of $\cL$. We say that a matrix $\mat{B}$ forms
a basis of $\cL$ if its rows form a basis of $\cL$.
All the bases have the same number $n$ of elements, called the \emph{dimension} or \emph{rank}
of $\cL$. 
The dual $\widehat{\cL}$ of a lattice $\cL$ is the set of all vectors $\vec{x}\in\operatorname{span}(\cL)$ such that $\Angle{\vec{x},\vec{y}}$ is an integer for all $\vec{y}\in \cL$.  If $\mat{B}$ is a basis of $\cL$ then $\widehat{\mat{B}}\coloneqq {(\mat{B}\cdot \mat{B}^{T})}^{-1}\cdot \mat{B}$
is a basis of $\widehat{\cL}$.  We call  $\widehat{\mat{B}}$ the ``dual basis'' of $\mat{B}$. Lattice algorithms often involve the orthogonal projections \(\pi_{i}: \mathbb{R}^{n} \mapsto {\operatorname{span}\left(\vec{b}_1, \ldots, \vec{b}_{i-1}\right)}^{\perp}\) for \(i=1,\ldots,n\). The \emph{Gram--Schmidt orthogonalisation} (GSO) of \(\mat{B}\) is \(\mat{B}^{\ast} = (\vec{b}_1^{\ast}, \ldots, \vec{b}_{n}^{\ast})\), where the Gram--Schmidt vector \(\vec{b}_{i}^{\ast}\) is \(\pi_{i}(\vec{b}_{i})\).



One of the fundamental algorithmic problems related to lattices is to find a shortest non-zero element of an arbitrary lattice (with respect to its Euclidean norm), given an arbitrary basis of this lattice. This problem is referred to as the shortest vector problem ($\SVP$) and the length of such a vector denoted by $\lambda_1(\cL)$. It is a central premise of lattice-based cryptography  that solving \(\SVP\) (and its decision variant \textsf{GapSVP}) within a polynomial factor takes super-polynomial time also on a quantum computer~\cite{STOC:Regev05}. It is well-known that the hardness of the \SVP{} is related to the ``quality'' of the input basis which informally quantifies the length of the vectors in the basis and the angles between. Intuitively, a basis with short and relatively orthogonal vectors is of higher quality. Therefore, a fundamental problem in lattice-based cryptography is to increase the quality of a given basis, a process known as lattice reduction. The celebrated LLL algorithm~\cite{LLL82} was the first polynomial-time algorithm that computes a reduced basis of guaranteed quality, namely the first vector is at most exponentially longer than the shortest vector of the lattice. The BKZ algorithm~\cite{Schnorr1987AHO} is a generalisation of LLL to obtain more strongly reduced basis at the expense of a higher running time. More precisely, the BKZ algorithm requires us to choose a so-called block size $\beta$: the larger the $\beta$, the stronger the reduction but the higher the running time (which is at least exponential in $\beta$). BKZ internally uses an algorithm to solve (near) exact $\SVP$ in lower-dimensional lattices. Therefore, understanding the complexity of $\SVP$ is critical to understanding the complexity of BKZ and lattice reduction. This, in turns, is critical to choosing security parameters of cryptographic primitives~\cite{ACDDPPVW18}.

When \(\beta = n\) we recover the HKZ algorithm producing HKZ reduced bases, which one of the strongest notion of reduction~\cite{Kannan83}. The first vector of such a basis is always a shortest vector of the lattice. Furthermore, HKZ basis naturally lend themselves to be computed recursively and enjoy many good properties (see \cref{sec:hkz_reduction} for more details), especially in conjunction with enumeration~\cite{HS07}. Enumeration algorithms list all of the lattice points within a ball of prescribed radius $r$. One of the most important aspect of enumeration is to correctly chose the enumeration radius $r$ so that it is larger than $\lambda_1(\cL)$, but not too large since the running time increases rapidly with $r$. For random lattices, the so-called Gaussian Heuristic gives a good estimate of $\lambda_1(\cL)$ as $ \operatorname{gh}(\cL)\coloneqq \sqrt{\frac{n}{2\pi e}}\cdot{\det(\cL)}^{1/n}$.

The fastest known (heuristic) quantum algorithm~\cite{EPRINT:ChaLoy21} for solving \(\SVP\) is a ``sieving algorithm'' runs in  time \(2^{0.257\,n+o(n)}\), uses QRAM~\cite{GLM08} of maximum size \(2^{0.0767\,
n+o(n)}\), a quantum memory of size \(2^{0.0495\,n+o(n)}\) and a classical memory of size \( 2^{0.2075\,n+o(n)}\).
The second main class of quantum algorithms for solving \(\SVP\) are ``lattice-point enumeration'' algorithms which combine exhaustive search with projections~\cite{AC:AonNguShe18}. These algorithms run in time \(n^{n/(4e) + o(n)}\) and \(\poly[d]\) memory. In many cryptography applications \(n^{n/16  + o(n)}\) seems plausible~\cite{C:ABFKSW20}. On classical computers, the current record computation solved the ``Darmstadt SVP Challenge'', which asks to solve a slightly relaxed version of \SVP{}, in dimension 180~\cite{EC:DucSteWoe21} using sieving on GPUs. Both classes of algorithms rely on Grover's algorithm or random walks which require long running computations on fault-tolerant quantum computers and thus are not suitable on quantum devices of the next decade, the NISQ era.


\paragraph{Variational Quantum Algorithms.} (VQA)s~\cite{vqaSurvey} are hybrid classical-quantum algorithms that can solve a variety of problems, including optimisation problems. VQAs are believed to be one of the most promising approaches to use and possibly offer quantum advantage on near-term quantum devices -- also known as Noisy Intermediate-Scale Quantum (NISQ) devices. The quantum devices that currently exist and those that will become available in the coming 2-5 years are expected to have at most 1000 qubits (intermediate-scale) and have imperfections (noisy). Since the number of qubits is limited, to run computations of interest one cannot ``afford'' to perform quantum error-correction since this would require an overhead in the order of $\approx 10^3$ physical qubits for every useful logical qubit. VQAs mitigate the effects of noise by a different approach. A computationally expensive subroutine is solved by the quantum device that amounts in estimating the ``energy'' of a quantum state that arises as the output of a parameterised quantum circuit of short depth (avoiding the excessive accumulation of errors). The preparation and measurement is run multiple times and the output is fed to a classical optimisation part that essentially off-loads part of the computation to a classical device. VQAs appear to offer advantages over classical algorithms in various areas of quantum chemistry~\cite{qchemistry1,qchemistry2} and is a promising approach  for many other areas including combinatorial optimization~\cite{isingFormulations}, quantum machine learning~\cite{metaVariationalVQE} and quantum metrology~\cite{quantumMetrology}.

Since SVP can be formulated as an optimisation problem, we focus on two most widely used VQAs for combinatorial optimisation, namely Variational Quantum Eigensolver (VQE)~\cite{vqe1,vqe2} and Quantum Approximate Optimisation Algorithm (QAOA)~\cite{qaoa}. The first step for both of them, as well as for quantum annealing~\cite{quantum_annealing}, is to encode the problem (SVP here) to the ground state (smallest eigenvalue) of a Hamiltonian operator $\mathcal{H}$. For QAOA and quantum annealing, the Hamiltonian needs to be an Ising spin Hamiltonian (i.e.~involving only $Z$ spins and up to quadratic interaction terms) and for quantum annealing extra limitations due to the connectivity of the spins apply. For VQE the Hamiltonian can take more general form (including higher order terms and/or $X$ spins). The second step, once the Hamiltonian is chosen, is to select an \emph{ansatz state} $\ket{\psi(\theta)}$ -- a family of quantum states, parameterised by $\theta$, that are the output of a simple parameterised circuit. Ideally, we would like to ensure that elements of the family considered are close to the ground state of the problem's Hamiltonian. The two classes of ansätze that exist are the \emph{hardware efficient} ones that are essentially chosen for the ease that can be implemented at a given quantum hardware, and the \emph{problem specific} that are ansätze that use information about the problem for example using the problem's Hamiltonian. The former are less prone to errors and can be used with any Hamiltonian but have no guarantee to be ``dense'' around the true ground state, while the latter are more sensitive to noise and can be used with specific classes of Hamiltonians but are designed to have states close to the true ground state. The third step is to prepare a state from the ansatz and measure it where this step is repeated multiple times. From these repetitions an estimate of  the expectation $\bra{\psi(\theta)}\mathcal{H}\ket{\psi(\theta)}$ is calculated and passed as the cost value for the choice of parameters to a classical optimiser.

The optimiser then calculates new parameter $\theta$ and the procedure repeats until some stopping criterion is reached and an estimate for the ground state is produced. For our VQE runs we used hardware efficient ansätze. QAOA is by definition a problem specific ansatz since the family is constructed as  a discretised version of Quantum Adiabatic Computation. This puts a constraint on the form of the Hamiltonian (which makes it harder to solve $\SVP$ with fewer qubits see \cref{sec:encodingZeroConstraint}), but has the theoretical guarantee that for sufficiently deep circuits the solution should be found.
Specifically, for Hamiltonians that involve only spin $Z$ terms, to compute the energy/cost $C(\theta)$ of a quantum state $\ket{\psi(\theta)}$ we prepare the state and measure  in the computational basis $N$ times. Each run gives outcomes (bit-strings) $x_i$, and for every outcome we compute the related cost $m_i(x_i)$. Our estimate of the cost of the state is $C(\theta)=\bra{\psi(\theta)}\mathcal{H}\ket{\psi(\theta)}\approx\frac{1}{N}\sum_{i=1}^N m_i$ and is used by the classical optimiser. For classical combinatorial optimization problems one can find that other methods are performing better. The Conditional Vale at Risk (CVaR) \cite{cvar} and the Ascending-CVaR \cite{ascending_cvar} are two methods that give better results and in this work we will use the former. In \cite{cvar} instead of computing the cost taking the average of the values $m_i$, they considered ordering the values from the smallest to the larger and counting the $\alpha$-tail of the distribution. Specifically, $\alpha$ is to be chosen heuristically and the cost is calculated as an average of $\ceil{\alpha N}$ lowest measurements outcomes. Suppose $\{\tilde{m}_i\}_{i=1,...,N}$ is a sorted set of  $\{m_i\}_{i=1,...,N}$ in non-decreasing order. Then the cost is calculated as: $C_{CVaR_\alpha}(\theta)=\frac{1}{\ceil{\alpha N}}\sum_{i=1}^{\ceil{\alpha N}} \tilde{m}_i$.

We note that finding a ground state of a Hamiltonian is QMA-complete in general, but for specific Hamiltonians the ground \emph{can} be found efficiently -- adiabatic quantum computing is a universal model, that finds efficiently the ground state of those Hamiltonians that correspond to BQP problems. Indeed, in our case, since SVP is believed to be outside BQP we do not expect to find the solution efficiently. On the other hand, obtaining a polynomial speed-up is valuable for the cryptanalysis -- after all Grover's algorithm also provides such a moderate speed-up. In particular, we do not expect to do better than Grover -- query complexity bounds indicate that for our problem we can get at most a small constant improvement to Grover. However, due to the fact that our approach is heuristic, we may still be able to get considerable (larger) speed-up for certain (but not the hardest) instances. To properly analyse the time-complexity of a variational quantum algorithm, one needs to bound the scaling of the probability that the algorithm (at the end of the classical-quantum iterations) returns the correct solution – viewed differently, to bound the overlap of the output quantum state with the true shortest-vector. Then by repeating the algorithm sufficient\footnote{Here sufficient scales inversely with the probability of success of a single run} number of times, one is guaranteed to find the correct solution with high probability. While such bounds have been found for certain problems \cite{ksat,qaoa,qaoaApplied}, in the general case it is hard to obtain them (or even impossible), so this is left for future work. Note that the classical time-complexity scales exponentially, therefore even if this probability vanishes exponentially fast (as expected) our solution may still give competitive results offering similar performance to other ``fault-tolerant'' quantum algorithms.


%% file: relatedWork.tex
There have been several works focusing on translating \SVP{} into Hamiltonian $\mathcal{H}$ where the ground state corresponds to the shortest lattice vector.  Note that we can trivially achieve encodings where the corresponding eigenvalues of $\mathcal{H}$ define the order of lattice vector lengths. The resulting variational quantum formulation is thus capable of solving the approximate \SVP{}, a relaxation of \SVP{}, on which the security of most lattice-based protocols is relied upon.
In~\cite{notSoAdiabatic} the energy gaps between the first three excited states of the problem Hamiltonian are analysed when solving low dimensional \SVP{} via adiabatic computation. The ground state is by their construction a zero state and hence the first excited state is sought. The results suggest the existence of ``\textit{Goldilocks}'' zones. In such zones the adiabatic evolutions are slow enough that there is a high probability of sampling any of the first three excited states without any strong dominance of any of them. As a consequence, in this case it is possible to obtain the shortest non-zero vector with a small number of measurement samples. This motivates the use of QAOA to find the ground state as it mimics the adiabatic computation. However, the experiments were performed only for lattices up to 4 dimensions. Moreover, as \SVP{} is an integer optimisation problem, bounds on the ranges of the (integer) variables need to be defined prior to mapping the problem into a binary optimisation problem. The authors make an experimental guess that each of them grows linearly with lattice dimension resulting in guess of $O(n \log n)$ qubit requirement. The qubit requirement was later proved in~\cite{JCLM21} for special lattices with what they called ``an optimal Hermite Normal Form (HNF)''.  The density of such lattices is around $44\%$ for lattices generated with entries selected ``uniformly at random''~\cite[Section 5.1]{Maze10}.
For such lattices, they show that $\tfrac{3}{2}n\log_2 n +n+ \log \left(\det(\mathcal{L})\right)$ qubits are enough to guarantee that the shortest vector is an eigenvector of the hamiltonian.
To confirm the approach, experiments for up to 7-dimensional instances of \SVP{} were performed on D-Wave quantum annealer making use of up to 56 logical qubits.  The proof of their bound crucially relies on the special shape of
the HNF namely that it has at most one nontrivial column.
However,  $q$-ary lattices (which are ubiquitous in cryptography,
see \Cref{sec:exp_framework}) almost never have such a special Hermite Normal Form
since\footnote{Indeed, the matrix of a $q$-ary lattice is $\begin{bmatrix}\mat{I}_{n-k}&\mat{X}\\0&q\mat{I}_{k}\end{bmatrix}$, where $\mat{X}\in\mathbb{Z}_q^{(n-k)\times k}$
has rank $k$, is already in HNF.} they have $k$ nontrivial colums, where $k$ is the rank of the underlying
linear code and typically linear in the dimension (such as $n/2$).


%% file: mappingSVPtoHamiltonian.tex
Recall that the starting point for VQE and QAOA (but also for quantum annealing) is to encode the problem to the ground state of a Hamiltonian $\mathcal{H}$. As in other combinatorial optimisation problems, one first needs to turn the problem to a quadratic unconstrained binary (QUBO) form and then there is a standard method to obtain an Ising Hamiltonian (i.e. a Hamiltonian that involves only $Z$ operators and the interaction terms involve just pairs of qubits). Note that to run VQE one can have more general Hamiltonians and as we will see later, in some cases it is simple to include constraints (or even higher order terms). Coming back to QUBO formulatons the general form of the cost is  $C(s_1s_2...s_n)=c+\sum_i c_{ii}s_i + \sum_{i\neq j} c_{ij}s_is_j$, for string binary variables $s_1s_2\cdots s_n$, for some coefficients $c,\{c_{ij}\}_{1\leq i,j\leq n}$. Since Pauli-Z operators have $\pm 1$ eigenvalues one maps each binary variable $s_i$ to $\frac{I_i-Z_i}{2}$ leading to the Ising Hamiltonian,
$H=\sum_i c_{ii} \frac{I_i-Z_i}{2} + \sum_{i\neq j} c_{ij} \frac{I_i-Z_i}{2}\otimes\frac{I_j-Z_j}{2}$
where $Z_i$ and $I_i$ are the Pauli-Z and identity operators respectively, acting on $i$-th qubit.

\subsection{Mapping the Shortest Vector Problem into QUBO formulation}\label{sec:mappingSVintoQUBO}
A first approach (also starting point of~\cite{notSoAdiabatic,JCLM21}), is to consider as variables the coefficients of the basis vectors. Each choice of the coefficients determines a vector in the lattice, and its length can be easily calculated using the basis matrix. There are two major challenges with this approach. The first one is that the coefficients are integers while we want to use a finite number of binary variables. Recall that each binary variable will result to a qubit so we can only use a finite number while making this number as small as possible is crucial to run it in a NISQ device. We therefore need to truncate (limit) the possible coefficients, but in a way that the shortest vector is included in our possible solutions. The second challenge is that if the variables are the coefficients then a possible vector is the all zero vector, a vector that clearly is shorter than the shortest (non-zero) vector we are searching. We would therefore need to impose the constraint $x\neq 0$ at some level.

Going back to the formulation, given an $n$-dimensional full-rank row-major lattice basis matrix $B$, we define lattice $\mathcal{L}(\mat{B})=\{\vec{x} \cdot \mat{B} |\vec{x}\text{ row vector}, \vec{x}\in\mathbb{Z}^n\}$. The shortest vector problem ($SVP$) finds the solution
$	\lambda_1 \coloneqq \min_{\vec{y}\in\mathcal{L}(\mat{B})\setminus\{\vec{0}\}}\|\vec{y}\|$.
Let $\vec{x}$ be a row vector of coefficients and $\mat{G}=\mat{B}\cdot \mat{B}^T$ a Gram matrix of the basis. Then $y=\vec{x}\cdot \mat{B} \implies \|\vec{y}\|^2=\vec{x} \cdot \mat{B}\cdot \mat{B}^T\cdot \vec{x}^T=\vec{x}\cdot \mat{G} \cdot \vec{x}^T$ which allows us to reformulate it 
as a quadratic constrained integer optimisation problem: 
\begin{align}\label{eq:svpformulation2}
  \lambda_1^2&=\min_{\vec{y}\in\mathcal{L}(\mat{B})\setminus\{0\}}\|\vec{y}\|^2\\
  &=\min_{\vec{x}\in\mathbb{Z}^n\setminus\{\vec{0}\}}\sum_{i=1}^nx_i\cdot \mat{G}_{ii}+ 2 \sum_{1\leq i<j\leq n}x_i\cdot x_j \cdot \mat{G}_{ij}.\nonumber
\end{align}
As a consequence of rank-nullity theorem,  we have $\lambda_1\neq 0$ as required.  In order to convert (\ref{eq:svpformulation2}) into a binary optimisation problem, we need bounds $|x_i|\leq a_i$ for all $i=1,...,n$. This has been the core problem of all approaches to map the SVP into a Hamiltonian and we provide tighter and more general bounds in Section \ref{sec:compositeSVP}.

We now define new binary variables ${\{\tilde{x}_{ij}\}}_{0\leq j \leq \lfloor\log 2a\rfloor}$ using the initial integer variables $x_i$ and the bound $a_i$

{\small\begin{align}\label{eq:conversion_simple}
  & |x_i|\leq a\implies \\
  & x_i=-a+\sum_{j=0}^{\lfloor\log 2a\rfloor-1}2^j\, \tilde{x}_{ij}+(2a+1-2^{\lfloor\log 2a\rfloor})\cdot \tilde{x}_{i,\lfloor\log 2a\rfloor}\nonumber
\end{align}}
where the last term ensures we do not enlarge the search space. By substituting into (\ref{eq:svpformulation2}) and ignoring the  $\vec{x}\neq 0$ constraint, the resulting QUBO becomes (\ref{eq:qubo_ignore}) where $c$, $c_{ij}$ and $d_{ij,kl}$ are calculated constants resulting from the substitution.
\begin{equation}\label{eq:qubo_ignore}
	\min_{\substack{\tilde{x}_{1,0},...,\tilde{x}_{1, \lfloor\log 2a_1\rfloor}\\
			\dots\\
			{\tilde{x}_{n,0},...,\tilde{x}_{n, \lfloor\log 2a_n\rfloor}}}}
	c+\sum_{\tilde{x}_{i,j}}c_{i,j}\cdot \tilde{x}_{i,j}+\sum_{\tilde{x}_{ij}, \tilde{x}_{k,\ell}} d_{i,j,k,\ell}\cdot \tilde{x}_{i,j}\cdot \tilde{x}_{k,\ell}
\end{equation}

\subsection{Encoding the $x\neq 0$ constraint}\label{sec:encodingZeroConstraint}

In deriving \cref{eq:qubo_ignore} we ignored the $\vec{x}\neq 0$ condition in the definition of SVP\@. The true minimum is zero and is obtained by setting all the integer variables $x_1,\ldots,x_n$ to zero. Instead we want to obtain the second smallest value. In other words, we are seeking the first excited state of the corresponding Hamiltonian, with the extra information that the ``true'' ground state (zero vector) is known. There are three ways to address this. First is to ignore the constraint, run the optimisation to find the ansatz state with greatest overlap with the zero vector, and hope that it has an (ideally large) overlap with the first excited state \cite{notSoAdiabatic}. Even if this approach succeeds in finding SVP for small dimensions it is unlikely to work well at large scales. The second solution is to exclude the zero vector by imposing it in a form of a constraint that ``penalises'' the zero vector. To map the problem back to a QUBO, however, requires introducing extra variables ($n-2$ in our case), making this approach less practical. Since this approach can work both for QAOA and quantum annealing where the last and more practical approach does not work,  we give the details of this mapping in Appendix \ref{app:avoidZero}.

The third approach, that we will analyse here and used in our numerical experiments, can be used in VQE. One directly targets the first excited state by modifying the classical loop of the VQE i.e. modifying the way that the cost is evaluated. In the general form, assuming that $\ket{\psi_0}$ is the ground state, instead of using the expression $C(\psi)=\bra{\psi}H\ket{\psi}$ for the cost, we use $C(\psi)=\bra{\psi}H\ket{\psi}\frac{1}{1-\vert\bra{\psi}\psi_0\rangle\vert^2}$, where the multiplicative factor is the inverse of the probability of not being in the ground state. This gives infinite cost to the ground state, and in general penalises states with greater overlap with the ground state. It is easy to see that this cost is minimised at the first excited state of the original Hamiltonian. Note that other approaches to find the first excited state exist (see for example \cite{Higgott2019}).

In our case the ground state is the zero vector. Let $\tilde{N}$ be the cardinality of $m_i$'s (measured bit-strings) that are non-zero. Then $\frac{N}{\tilde{N}}$ is the numerical estimate of the inverse of the probability not being the zero vector. Therefore we will be using the modified cost $C'(\theta)=\left(\frac{N}{\tilde{N}}\right)\frac{1}{N}\sum_{i=1}^{N} m_i$. Equivalently, this means that to compute the cost of a state, we disregard the zero-vector outcomes taking the average value over all the other outcomes. In the measurements of the final ansatz state, at the end of the optimisation, we output the measurement sample with the lowest non-zero energy. The advantage of this approach is that the quantum part (states and measurements) are identical to the unconstrained one, and the only difference is in the way that the measurement outcomes are used to assign a cost to different states in the classical optimisation part of the VQE\@.

\subsection{Handling approximate solutions}
As the dimension of the lattice increases, the probability that the algorithm converges to a shortest vector might become vanishingly small. Instead, the system might return a short-but-not-shortest vector. In this case, one approach would be to replace a vector of the basis by the newly obtained one, making the basis more reduced, and restart. This approach has been experimentally observed to work in~\cite{ZJLM22}. In the classical setting, the use of \emph{tightly controlled} approximate SVP oracles was shown to provide an exponential speed-up over enumeration with exact SVP oracles~\cite{C:ABLR21}, but here we likely do not get to choose the looseness of our oracles. In principle, if we could characterise the distribution of the output of the algorithm and quantify the expected size of the vector, we could estimate the number of times the algorithm needs to be re-run before obtaining a shortest vector with good probability.
We leave this analysis as future work since it is nontrivial to estimate the distribution
of the output of the algorithm.


%% file: compositeSVP.tex

In the previous section we gave a map of \SVP{} to a QUBO formulation that is suitable for VQE. This map, however, relies on bounds on each $|x_i|$. In this section, we obtain worst case bounds on the $|x_i|$ based on the orthogonality defect of the dual basis. We then give two applications of this result. First, we show how to obtain a recursive algorithm to compute a HKZ reduced basis and thus solving the \SVP{} in passing. This method gives us the best asymptotical bound on the number of qubits required. Second, we estimate the number of qubits required to directly solve the \SVP{} by reducing the dual basis with recursive or classical preprocessing and then applying the NISQ enumeration.


%% file: optimalBounds.tex
As we have seen in \Cref{sec:mappingSVintoQUBO}, we need to
choose a bound on each of the $|x_i|$ to obtain a finite
optimisation problem. This bound will then determine the number
of qubits required. In order to be sure to find a solution, we first 
derive a general bound on all the $x_i$ that correspond to lattice
points in a ball of radius $\radenum$. We then use the Gaussian heuristic
to choose the radius $\radenum$ so as to ensure that this ball contains
at least one nonzero vector for most lattices and therefore a shortest vector.

\begin{lemma}\label{lem:bound_xi_precise_dual}
    Let $x_1,\ldots,x_n$ be such that $\norm{x_1\cdot \vec{b}_1+\cdots+x_n\cdot \vec{b}_n}\leqslant A$,
    then for all $i=1,\ldots,n$ we have
    $|x_i|\leqslant \radenum \cdot \|\hat{\vec{b}}_i\|$ where
    $\hat{\vec{b}}_1,\ldots,\hat{\vec{b}}_n$
    are the rows of $\widehat{\mat{B}}$, a dual basis of
    $\mat{B}$ which is the matrix whose rows are $\vec{b}_1,\ldots,\vec{b}_n$.
\end{lemma}
\begin{proof}
    Let $\delta_{i,j}$ denote the Kronecker delta.
    Observe that $\Angle{\vec{b}_j,\hat{\vec{b}}_i}=\delta_{i,j}$ for all $i,j$. Indeed,
    $\Angle{\vec{b}_j,\hat{\vec{b}}_i}={(\mat{B}\cdot \widehat{\mat{B}}^{T})}_{i,j}={(\mat{I}_n)}_{i,j}=\delta_{i,j}$.
    Now let $\vec{v}=x_1\cdot \vec{b}_1+\cdots+x_n\cdot \vec{b}_n$ for some $x_1,\ldots,x_n\in \ZZ$ be such that $\norm{\vec{v}}\leqslant \radenum$. Then for any $i$, we have that
    $\left|\Angle{\vec{v},\hat{\vec{b}}_i}\right|=|x_i|$. But on ther other hand,
    $\left|\Angle{\vec{v},\hat{\vec{b}}_i}\right|\leqslant\norm{\vec{v}} \cdot \|\hat{\vec{b}}_i\|$ which proves the result. \qed{}
\end{proof}

If we now make use of the Gaussian heuristic to choose the radius \(\radenum\) and
take $\radenum \coloneqq \gh(\cL) = \sqrt{n/2\pi e}\cdot {\vol(\cL)}^{1/n}$.
The total number of qubits that we need will be
\begin{equation}\label{eq:nr_qubits_enum}
  N = \sum_{i=1}^n\left(\lfloor\log_2(2m_i)\rfloor+1\right) \leqslant 2\,n+\log_2\prod_{i=1}^n m_i
\end{equation}
where $m_i$ is the bound on the $|x_i|$. Using the bound of \Cref{lem:bound_xi_precise_dual}
we obtain
\begin{align*}
\prod_{i=1}^n m_i
&\leqslant {\left(\sqrt{\frac{n}{2\pi e}}\right)}^n \cdot \vol(\cL) \cdot
\prod_{i=1}^n\norm{\hat{\vec{b}}_i}\\
& ={\left(\sqrt{\frac{n}{2\pi e}}\right)}^n \cdot
\frac{\prod_{i=1}^n\norm{\hat{\vec{b}}_i}}{\vol(\widehat{\cL})}\\
&={\left(\sqrt{\frac{n}{2\pi e}}\right)}^n\cdot \delta(\widehat{\cL})
\end{align*}
where $\widehat{\cL}$ is the dual of $\cL=\cL(\vec{b}_1,\ldots,\vec{b}_n)$ and $\delta(\cdot)$ denotes the orthogonality defect. This suggest that the critical factor for the enumeration is the reduction of the dual of the lattice and not the lattice itself.

\begin{corollary}\label{cor:bound_qubits_dual}
	The number of qubits required for the enumeration on the basis $\mat{B}$,
	assuming the Gaussian heuristic is bounded by
	$
	2\,n+\log_2\left({\left(\frac{C^2\cdot n}{2\pi e}\right)}^{n/2} \cdot \delta(\widehat{\mat{B}})\right)
	$
	where $\delta(\cdot)$ denotes the orthogonality defect
	and $C$ the multiplicative factor used with the Gaussian heuristic\footnote{The
		enumeration is done with a radius of $C\cdot \gh(\cL)$.}.
\end{corollary}

\subsection{Bound on the number of qubits for producing a HKZ basis}\label{sec:hkz_reduction}

In this section, we show that our NISQ enumeration procedure can be used to HKZ reduce a basis in a recursive way using only $\frac{3}{2}n\log_2(n)+O(n)$ qubits, thus solving SVP in passing. This is asymptotically the same as the algorithm in~\cite{JCLM21}. However, our algorithm works for any lattice, whereas the algorithm in~\cite{JCLM21} only works for lattices that have what they called ``an optimal Hermite Normal Form''. See \Cref{sec:relatedwork} for more details.

Recall that the notion of HKZ-reduced basis can be defined inductively as follows.
Any vector $\vec{b}$ is a HKZ-reduced basis of $\cL(\vec{b})$.
A basis $\vec{B}=(\vec{b}_1,\ldots,\vec{b}_k)$, for $k\geqslant 2$, is HKZ-reduced if
\begin{itemize}
	\item $\vec{b}_1$ is a shortest vector in $\cL(\vec{B})$,
	\item $(\pi_2(\vec{b}_2),\ldots,\pi_2(\vec{b}_k))$ is HKZ-reduced, where \(\pi_{2}(\cdot)\) is the projection on $\vec{b}_1^\perp$,
	\item $|\Angle{\vec{b}_1,\vec{b}_i}|\leqslant\frac{1}{2}|\Angle{\vec{b}_1,\vec{b}_1}|$.
\end{itemize}
Intuitively, the first two conditions mean that $\vec{b}_1$ is a shortest vector of the lattice $L(\vec{b}_1,\ldots,\vec{b}_k)$, and after projecting on $\vec{b}_1^\perp$, $\pi_2(\vec{b}_2)$ is a shortest vector of the projected lattice $\cL(\pi_2(\vec{b}_2),\ldots,\pi_2(\vec{b}_k))$, and so on. This notion of reduction is much stronger than LLL-reduced basis and BKZ-reduced basis. The third condition is equivalent to the notion of size-reduction which is also required by LLL-reduced basis and is a technical condition.

We say that a basis $\vec{B}=(\vec{b}_1,\ldots,\vec{b}_n)$ is \emph{pseudo-HKZ-reduced} if it is an LLL basis on which we apply an HKZ reduction to the \emph{first} $n-1$ vectors. This intuitively means that the basis is almost HKZ-reduced, except for the last vector, and the last vector is still controlled by the fact it comes from an LLL-reduction. In particular, the last vector is guaranteed to be at most exponentially longer than a shortest vector of the lattice.

Our algorithm is a variant of the classical algorithm for producing a HKZ reduced basis due to Kannan~\cite{Kannan83}. The main difference is that Kannan's algorithm exclusively works on the (primal) basis whereas our algorithm has to reduce both the primal and the dual basis to control the number of qubits required in the enumeration steps. This difference requires us to work slightly differently. In particular, Kannan's algorithm works by recursively producing a ``quasi-HKZ'' reduced basis, and then applying enumeration on this basis to obtain a full HKZ basis. A \emph{quasi-HKZ reduced} basis is essentially a LLL basis on which we apply an HKZ reduction to the projection of the \emph{last} $n-1$ vectors (orthogonally to the first). In contrast, our algorithm first produces a pseudo-HKZ basis, i.e.~an LLL-reduced basis on which we apply an HKZ reduction to the \emph{first} $n-1$ vectors. Our algorithm then runs enumeration on the dual.

\begin{algorithm}
	\DontPrintSemicolon\SetAlgoVlined\SetAlgoNoEnd\LinesNumbered
	
	\SetKwInOut{Input}{input}\SetKwInOut{Output}{output}
	\Input{A basis $(\vec{b}_1,\ldots,\vec{b}_n)$}
	\Output{An HKZ basis of the same lattice}
	\BlankLine
	$\cL:=\cL(\vec{b}_1,\ldots,\vec{b}_n)$\;
	$(\vec{d}_1,\ldots,\vec{d}_n)$ := dual basis of $(\vec{b}_1,\ldots,\vec{b}_n)$\;\label{alg:dhq:dual_comp_1}
	$(\vec{d}_1,\ldots,\vec{d}_n)$ := LLL-reduce $(\vec{d}_1,\ldots,\vec{d}_n)$\;\label{alg:dhq:lll_red_1}
	$(\vec{d}_1,\ldots,\vec{d}_{n-1})$ := HKZ-reduce $(\vec{d}_1,\ldots,\vec{d}_{n-1})$
	\tcc*{the basis is $(n-1)$-dimensional} \label{alg:dhq:hkz_red_1}
	$(\vec{b}_1,\ldots,\vec{b}_n)$ := dual basis of $(\vec{d}_1,\ldots,\vec{d}_n)$\; \label{alg:dhq:dual_comp_2}
	Call the NISQ enumeration procedure proposed in \Cref{sec:svpWithVQA} on $(\vec{b}_1,\ldots,\vec{b}_n)$ using the bounds established in \Cref{lem:bound_xi_precise_dual} and keep a shortest vector $\vec{v}$\;\label{alg:dhq:enum}
	$(\vec{b}_1,\ldots,\vec{b}_n)$ := LLL-reduce $(\vec{v},\vec{b}_1,\ldots,\vec{b}_n)$
	\tcc*{extract a basis from $n+1$ vectors}\label{alg:dhq:lll_red_2}
	Compute the orthogonal projections $(\vec{b}_2',\ldots,\vec{b}_n')$ of $(\vec{b}_2,\ldots,\vec{b}_n)$
	on $\vec{b}_1^\perp$\;\label{alg:dhq:ortho_proj}
	$(\vec{b}_2',\ldots,\vec{b}_n')$ := HKZ-reduce $(\vec{b}_2',\ldots,\vec{b}_n')$
	\tcc*{the basis is $(n-1)$-dimensional}\label{alg:dhq:hkz_red_2}
	For each $i$, set $\vec{b}_i:=\vec{b}_i'+\alpha_i\vec{b}_1$ where $\alpha_i\in\left(-\tfrac{1}{2},\tfrac{1}{2}\right]$ is such that $\vec{b}_i$
	belongs to $L$\;\label{alg:dhq:lift}
	\Return{$(\vec{b}_1,\vec{b}_2,\ldots,\vec{b}_n)$}\;
	\caption{HKZ-reduction algorithm using quantum enumeration\label{alg:dual_hkz_quantum}}
\end{algorithm}

We will require the following well-known result about HKZ basis.

\begin{proposition}[\cite{WenC19}]\label{prop:ortho_defect_hkz}
	The orthogonality defect of a HKZ basis is bounded by
	$
	\gamma_n^{n/2} \cdot \prod_{i=1}^n\cdot \frac{\sqrt{i+3}}{2}
	$
	where $\gamma_n$ is Hermite’s constant in dimension $n$, and $\gamma_n<\tfrac{1}{8}n+\tfrac{6}{5}$.
\end{proposition}

\begin{corollary}\label{cor:ortho_defect_hkz}
	The orthogonality defect of a HKZ basis is $\leqslant 2^{n\log_2n-\left(\tfrac{5}{2}+\tfrac{1}{2}\log_2e\right)n+O(\log n)}$.
\end{corollary}
\begin{proof}
See \Cref{sec:proof_ortho_defect_hkz}.  
\end{proof}

\begin{theorem}\label{th:hkz_quantum_th}
	Assuming the Gaussian heuristic, Algorithm~\ref{alg:dual_hkz_quantum} produces an HKZ basis with $\frac{3}{2}n\log_2(n)-2.26n+O(\log n)$ qubits.
\end{theorem}
\begin{proof}
  We first show that the algorithm is correct by induction on $n$.
  Let $\cL=\cL(\vec{b}_1,\ldots,\vec{b}_n)$ at the beginning of the algorithm.
  First note that at the beginning of line~\ref{alg:dhq:enum}, the basis $(\vec{b}_1,\ldots,\vec{b}_n)$ is still a basis of the original lattice $\cL$ since all the previous operations preserve that property.
  Assuming the Gaussian heuristic is true (or true with a multiplicative factor\footnote{Many papers and the SVP challenge \cite{SVPchallenge}, typically use the GH with a multiplicative factor $C\approx 1.05$. For large $n$, Minkowski's bound on the shortest vector implies that $C=\sqrt{2\pi e}$ actually holds but this is a much worse bound.} $C$), the enumeration algorithm will indeed find a shortest nonzero vector $\vec{v}$ at line~\ref{alg:dhq:enum}.
  The reduction at line~\ref{alg:dhq:lll_red_2} will allow to extract a basis of $\cL$ among the $n+1$ vectors $\vec{v},\vec{b}_1,\ldots,\vec{b}_n$.
  Since $\vec{v}$ is a shortest vector, the LLL reduction will ensure that $\vec{b}_1$ is a shortest vector of $\cL$ at the beginning of line~\ref{alg:dhq:ortho_proj}.
  Let $\cL'=\cL(\vec{b}_2',\ldots,\vec{b}_n')$ be the projected lattice obtained after line~\ref{alg:dhq:ortho_proj}. The reduction at line~\ref{alg:dhq:hkz_red_2} does not change the lattice spanned by the $\vec{b}_i'$ which is now a HKZ basis. The lifting operation at line~\ref{alg:dhq:lift} ensures that $(\vec{b}_1,\ldots,\vec{b}_n)$ is a basis of $\cL$ at the end of the algorithm. Since $\vec{b}_1$ is a shortest vector of $L$ and $(\vec{b}_2',\ldots,\vec{b}_n')$ is HKZ and a basis of the projected lattice, it follows that the returned basis is HKZ\@.
	
  We now analyse the qubits requirement of the algorithm. Let $(\bar{\vec{d}}_1,\ldots,\bar{\vec{d}}_n)$ be the basis obtained after line~\ref{alg:dhq:lll_red_1} and $(\bar{\vec{d}}_1^*,\ldots,\bar{\vec{d}}_n^*)$ be its Gram-Schmidt orthogonalisation.
  By the properties of the LLL-reduction, we have $\norm{\bar{\vec{d}}_n}^2\leqslant 2^{n-1}\cdot \norm{\bar{\vec{d}}_n^*}^2$~\cite{NguVal10}. Let $(\vec{d}_1,\ldots,\vec{d}_n)$ be the basis obtained after line~\ref{alg:dhq:hkz_red_1} and $(\vec{d}_1^*,\ldots,\vec{d}_n^*)$ be its Gram-Schmidt orthogonalization. Clearly $\vec{d}_n=\bar{\vec{d}_n}$ since this vector was not touched by the HKZ reduction. It follows that $\vec{d}_n^*=\bar{\vec{d}}_n^*$ because $\vec{d}_n^*$ (resp.~$\bar{\vec{d}}_n^*$) is the projection of $\vec{d}_n=\bar{\vec{d}}_n$ on the orthogonal of $\Span(\vec{d}_1,\ldots,\vec{d}_n)=\Span(\bar{\vec{d}}_1,\ldots,\bar{\vec{d}}_n)$ which are equal because the HKZ reduction does not change the span. As a result, we have $\norm{\vec{d}_n}^2\leqslant 2^{n-1}\cdot \norm{\vec{d}_n^*}^2$ after line~\ref{alg:dhq:hkz_red_1} and therefore
	\begin{align*}
	\delta(\vec{d}_1,\ldots,\vec{d}_n)
	& =\delta(\vec{d}_1,\ldots,\vec{d}_{n-1}) \cdot
	\frac{\norm{\vec{d}_n}}{\norm{\vec{d}_n^*}}\\
	&= 2^{n\log_2n-\left(2+\tfrac{1}{2}\log_2e\right)n+O(\log n)}
	\end{align*}
	by \Cref{cor:ortho_defect_hkz}. Taking the dual of this basis at line~\ref{alg:dhq:dual_comp_2} means at line~\ref{alg:dhq:enum}, we are in a position to run the enumeration algorithm on a basis $(\vec{b}_1,\ldots,\vec{b}_n)$ whose dual $(\vec{d}_1,\ldots,\vec{d}_n)$ has a small orthogonality defect. By \Cref{cor:bound_qubits_dual}, the number of qubits required for this enumeration step is bounded by
	\begin{align*}
	& 2n+\log_2\left[{{\left(\frac{C^2n}{2\pi e}\right)}^{n/2}\delta(\vec{d}_1,\ldots,\vec{d}_n)}\right]\\
	& \leqslant \tfrac{3}{2}n\log_2n \\
    & - \left(\tfrac{1}{2}\log_2\pi+\log_2e-\log_2 C\right)n+O(\log n)
	\end{align*}
	Denote by $Q(n)$ the number of qubits necessary to run the algorithm in dimension $n$. Since we run the enumerations sequentially, $Q(n)=\max(Q(n-1),n\log_2(n)+O(n))$
	and therefore $Q(n)=\frac{3}{2}n\log_2(n)-2.26n+O(\log n)$
	for $C=1$. \qed{}
\end{proof}

\subsection{Solving the SVP directly in the NISQ era}
As we have seen in the previous sections, the bound on the $x_i$ and thus the running time of the enumeration fundamentally depend on the quality of the dual basis. The algorithm of the previous section takes advantage of this fact by recursively reducing the primal and dual basis and always running the enumeration on a dual quasi-HKZ reduced basis, one of the strongest notion of basis reduction. Unfortunately, the bound on the number of qubits that we obtained relies on worst case bounds on the orthogonality defect of such basis. In particular, the linear term in the bound of \Cref{th:hkz_quantum_th} is quite pessimistic for most bases.

In this section, we perform numerical experiments to understand the number of qubits necessary to run the quantum enumeration depending on the quality of the dual basis. As mentioned above, the cost of enumeration is affected by the quality of the input basis. We thus consider input bases preprocessed with LLL or BKZ-\(\beta\). Using \Cref{cor:bound_qubits_dual}, we draw Figure~\ref{fig:qubits_with_dim} which shows the number of qubits needed when the dual basis is LLL reduced or BKZ-$\beta$ reduced with different $\beta$.  We choose the maximum value of $\beta=70$, since the running time becomes prohibitively long for larger $\beta$.
We also performed experiments with dual pseudo-HKZ reduced basis to understand the practical behavior of the algorithm in the previous section\footnote{Given the high cost of computing HKZ reduced basis, our experiments are limited to dimension 80.}.
The graph was obtained by generating random
$q$-ary lattices (for $q=65537$ and $k=n/2$),
reducing them and then computing $N$ as in \eqref{eq:nr_qubits_enum} using the bounds in \Cref{lem:bound_xi_precise_dual}. Each experiment was repeated
5 times\footnote{The q-ary lattices that we generate have fixed determinant, hence the number of qubits only depends on the orthogonality defect. We observed that the orthogonality defect of LLL and BKZ reduced basis varies little between runs, hence the small number of runs.} and the average was taken. We also performed regression using quadratic polynomials for LLL and BKZ reduced basis. This type of regression is expected to fit well since the number of qubits can be shown to be quadratic. On the other hand, as shown in \Cref{th:hkz_quantum_th}, for pseudo-HKZ reduced dual basis, the number of qubits should grow as $\Theta(n\log_2(n))$ asymptotically.  Hence, we fitted with a curve of form $\tfrac{3}{2}n\log_2(n)+an+b$.

In order to solve the current lattice approximate SVP record (dimension 180), we need approximately 1157 qubits when the dual is sufficiently reduced  ($\beta=70$ for example). This is considered to be achievable in the not so far future. We stress, however, that this does not imply such an algorithm will converge on a good solution efficiently but only that it can be run.

\input{fig_experimental_results_new.tex}


%% file: fig_experimental_results_new.tex
\begin{figure}
	\caption{Experimental results}\label{fig:experimental-results}
	\begin{subfigure}[t]{0.5\textwidth}
		\begin{tikzpicture}[scale = 1]
			\begin{axis}[xlabel={$n$},ylabel={Qubits},
				legend columns=1,
				width=1.0\textwidth,height=1.0\textwidth,,
				minor y tick num=0,
				xtick distance=20,
				ytick distance=300,
				minor x tick num=0,
				enlarge x limits=0.05,
				enlarge y limits=0.02,
				legend style={legend pos=north west},
				legend cell align={left},
				domain=20:150]
				\addplot[blue, only marks, mark size=1.5pt]
				table[x=x,y=y,y error plus=y+, y error minus=y-]
				{qubits_qary-dual-lll-gh.txt};
				\addplot[blue,domain=20:180,forget plot] {-21.699 + 2.467*x + 0.036*x^2};
				\addlegendentry{\textattachfile[color=.9 .2 .2]{qubits_qary-dual-lll-gh.txt}{LLL}};
				
				\addplot[green!70!black, only marks, mark size=1.5pt]
				table[x=x,y=y,y error plus=y+, y error minus=y-]
				{qubits_qary-dual-bkz-20-gh.txt};
				\addplot[green!70!black,domain=20:190,forget plot] {-37.785 + 3.221*x + 0.023*x^2};
				\addlegendentry{\textattachfile[color=.9 .2 .2]{qubits_qary-dual-bkz-20-gh.txt}{BKZ-20}};
				
				\addplot[red, only marks, mark size=1.5pt]
				table[x=x,y=y,y error plus=y+, y error minus=y-]
				{qubits_qary-dual-bkz-50-gh.txt};
				\addplot[red,domain=20:190,forget plot] {-16.164 + 2.673*x + 0.024*x^2};
				\addlegendentry{\textattachfile[color=.9 .2 .2]{qubits_qary-dual-bkz-50-gh.txt}{BKZ-50}};
				
				\addplot[brown, only marks, mark size=1.5pt]
				table[x=x,y=y,y error plus=y+, y error minus=y-]
				{qubits_qary-dual-bkz-70-gh.txt};
				\addplot[yellow!80!black,domain=20:190,forget plot] {-34.389 + 3.332*x + 0.018*x^2};
				\addlegendentry{\textattachfile[color=.9 .2 .2]{qubits_qary-dual-bkz-70-gh.txt}{BKZ-70}};
				
				\addplot[violet, only marks, mark size=1.5pt]
				table[x=x,y=y,y error plus=y+, y error minus=y-]
				{qubits_qary-dual-pseudohkz-gh.txt};
				\addplot[violet,domain=20:190,forget plot] {44.078 + -5.556*x + 3*x*log2(x)/2};
				\addlegendentry{\textattachfile[color=.9 .2 .2]{qubits_qary-dual-pseudohkz-gh.txt}{pseudo-HKZ}};
				
				\addplot[teal,domain=20:190] {-2.26*x + 3*x*log2(x)/2};
				\addlegendentry{\Cref{th:hkz_quantum_th}};
				
			\end{axis}
		\end{tikzpicture}
		\caption{Average number of qubits as a function of the dimension and their regression; input lattices are \(q\)-ary; reduction performed the dual lattice;  ``\Cref{th:hkz_quantum_th}'' is worst case bound with $O(\log(n))$ term omitted; raw data linked in red.\label{fig:qubits_with_dim}}
	\end{subfigure}\hspace{3mm}
	\begin{subfigure}[t]{0.5\textwidth}
		\begin{tikzpicture}
			\begin{groupplot}[
				group style={
					group name=my plots,
					group size=1 by 2,
					ylabels at=edge left,
					vertical sep=1cm
				},
				footnotesize,
				width=\textwidth,
				height=0.53\textwidth,
				tickpos=left,
				enlarge x limits=false
				]
				\nextgroupplot[title=,xlabel=\(n\), ylabel=Overlap, 
				ylabel shift=-1em,
				yticklabel style={
					/pgf/number format/fixed,
					/pgf/number format/precision=4,
				},
				scaled y ticks=false,
				xmin=14,xmax=29,
				ymin=-0.05,ymax=0.15]
				\addplot[blue, solid, mark=o]
				plot [error bars/.cd, y dir = both, y explicit]
				table[x =x, y =y, y error =e]{experimental_results/quantum_exp_data/hit_rates.txt};
				
				\nextgroupplot[title=,xlabel=\(n\), ylabel=Iterations,
				xmin=14,xmax=29]
				\addplot[blue, solid, mark=o]
				plot [error bars/.cd, y dir = both, y explicit]
				table[x =x, y =y, y error =e]{experimental_results/quantum_exp_data/convergence_times.txt};
			\end{groupplot}
		\end{tikzpicture}
		\caption{Top: Averaged overlap(inner-product) of state resulting from running the final ansatz circuit with Hamiltonian's 1st excited state (ground state is the zero vector), Bottom: Averaged number of VQE iterations until convergence}\label{fig:expResults}
	\end{subfigure}
\end{figure}
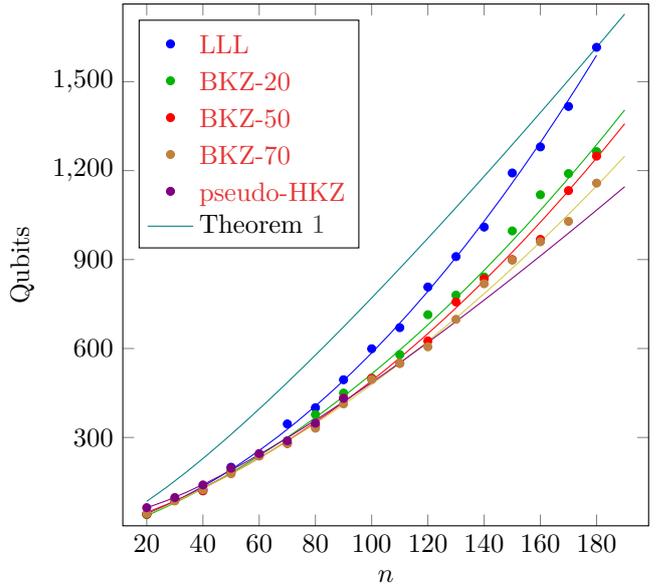

%% file: experimental_results.tex
We run quantum emulation of the VQE algorithm on a classical computer to assess the performance of our SVP solver. Due to the limited number of emulable qubits we considered lattices with rank up to 28, much smaller than the cryptographically relevant ranks but much bigger than prior quantum attempts. While we estimated the space-complexity required to solve large dimensions, we cannot reliably extrapolate the performance (accuracy/time) due to the heuristic nature of VQE. The effects of noise and error-mitigation are beyond the scope of this work and are left for future explorations.

\subsection{Experimental framework}\label{sec:exp_framework}

For our experiments we first sample
\[
  \mat{A} \coloneqq  \begin{pmatrix}
              \mat{I}_{d-k} & \mat{\widetilde{A}}\\
              \mat{0} & q\cdot \mat{I}_k
             \end{pmatrix}  \in \ZZ^{d \times d}
\]
where \(\mat{\widetilde{A}} \sample \ZZ_{q}^{(d-k) \times k}\) and \(\mat{I}_{x} \in \ZZ^{x \times x}\) is the identity. We then consider the lattice \(\cL(\mat{A})\) spanned by the rows of this matrix. In particular, we first LLL reduce the entire lattice basis and then consider the sublattice of rank \(n\) spanned by the first rows of the reduced basis. Our choice of \(q\)-ary lattices \(\cL(\mat{A})\) is partly motivated by their ubiquity in cryptography and by their connection to worst-case lattice problems. For example, finding short vectors in the dual of a random \(q\)-ary lattice is as hard as finding short vectors in any lattice~\cite{STOC:Ajtai96}. We preprocess these lattices using the polynomial-time LLL algorithm for numerical stability reasons. We extract a \(n\)-rank sublattice in dimension \(d\) rather than considering full-rank lattices with \(n=d\) since LLL will succeed in solving SVP for such small dimensions directly. The VQE receives a \(n \times d\) matrix as input and this ``leaves some work'' for the quantum algorithm to do. Note that even though many instances of these sub-lattices were ``almost'' solved by the described procedure, it is little advantage for the VQE as the size of the search space was not reduced. Therefore, by using a random guess for initial optimization parameters of an ansatz circuit, our experiments represent scaled-down real instances. Consequently, the hardness of SVP for VQE is in our experiments not even lowered if the shortest vector is already present in the given basis. The search space is defined by bounds on the coefficient vector as discussed in \cref{sec:optBound}. However, the bounds there are worst-case and asymptotic. In contrast, lattice algorithms tend to perform much better on average than in the worst case, especially in small dimensions. Indeed, \cref{fig:qubits_with_dim} illustrates the same phenomenon here.
We hence present a different qubit mapping strategy for which we compute probabilities of the shortest lattice vector lying within these bounds and in our quantum experiments we evaluate the probability of finding the ground state of problem Hamiltonian. The advantage is that the two problems: choosing an appropriate qubit mapping and finding the ground state of the problem Hamiltonian, can be tackled separately. The overall success probability of the variational SVP solver is then a product of the probability of encoding the shortest non-zero lattice vector in the search space and the probability of finding the Hamiltonian's ground state.

\paragraph{Naive qubit mapping approach.} Given an $n$-rank sublattice we assign one qubit per coefficient of the coefficient vector, i.e. $|x_i|\leq 1$ for $1\leq i \leq n$. Using this approach we found that the shortest non-zero lattice vector was included in the search space with the following probabilities: \textit{rank 15}: 80\%,  \textit{rank 16}: 75\%, \textit{rank 17}: 75\%, \textit{rank 18}: 74\%, \textit{rank 19}: 71\%, \textit{rank 20}: 70\%, \textit{rank 21}: 64\%, \textit{rank 22}: 62\%, \textit{rank 23}: 57\%, \textit{rank 24}: 55\%, \textit{rank 25}: 50\%. See \cref{app:groundStateProbs} for analysis of naive qubit mappings with ranks up to 50 that also considers different naive mapping strategies. The results presented in the appendix can be of an interest for all experimental approaches to the $SVP$ that use a Hamiltonian formulation.

\paragraph{Quantum emulation setup.}
We developed distributed variational quantum algorithms emulation software framework\footnote{\url{https://github.com/Milos9304/FastVQA}, \url{https://github.com/Milos9304/LattiQ}} by modification of parts of \textit{Xacc} \cite{xacc_2020} library to utilize implementations of variational quantum algorithms and useful classical optimisers and we use \textit{QuEST} \cite{quest} for the underlying quantum emulation. The experiments have been run on Ngio 5, a distributed computer provided by Edinburgh Parallel Computing Centre featuring 2x Xeon Platinum 8260 24C 2.4GHz with 12x 16GB DRAM modules. The emulations use state-vector representation of intermediate quantum states and do not consider any effects of noise. Noise would affect the probability of finding the ground state (uncertainty of cost makes the classical optimization harder) (\cref{fig:expResults} Top), but it would not affect the time requirements (\cref{fig:expResults} Bottom), neither the number of qubits. As it has been noted, extrapolating our results to lattice dimensions of cryptographic interest is infeasible from our experiments, because the way the probability of success decays cannot be extrapolated (among other reasons because we cannot anticipate the effects that barren plateaus would have). This is why for our current contribution adding noise would not add much in our analysis, beyond perhaps, bringing lower the probabilities of solving the problem in the very low qubit experiments that are possible to emulate.

\subsection{Experimental results}
To improve the performance in our experiments we used $CVaR_\alpha$ cost for VQE\@. We chose $\alpha=0.175$ since this gave better results (see also \cref{app:conditionalValAtRisk}). We were able to solve the SVP with our emulator for lattices with rank up to 28. Our experiments with 128 instances suggest that the success probability of finding the shortest non-zero lattice vector remains roughly constant for lattice instances with ranks not much larger than 28. \Cref{fig:expResults} (top) depicts the averaged overlap of the final ansatz state with the ground state corresponding to the shortest non-zero lattice vector found by a classical enumeration. The overlap represents the probability of sampling the shortest lattice vector with a single measurement of the final ansatz state. From the figure we see that we need $\approx 25$ samples to obtain the solution. Moreover, we observed linear time scaling (see \cref{fig:expResults} bottom). We note, however, that the ranks of cryptographically relevant lattices are larger ($\approx 400$) and we cannot extrapolate our observations with confidence.



%% file: acknowledgements.tex
The authors acknowledge support from EPCC, including use of the NEXTGenIO system, which was funded by the European Union's Horizon 2020 Research and Innovation programme under Grant Agreement no. 671951.  The research of MA was supported by EPSRC grants EP/S020330/1, EP/S02087X/1 and by the European Union Horizon 2020 Research and Innovation Program Grant 780701. The bulk of this work was done while MA was at Royal Holloway, University of London. MP acknowledges support by EPSRC DTP studentship grant EP/T517811/1. The research of YS was supported by EPSRC grant EP/S02087X/1 and EP/W02778X/1. The research of PW was supported by EPSRC grants EP/T001062/1, EP/T026715/1 and by the ISCF grant 10001712.

%% file: appendix_zeroVect.tex
Section \ref{sec:mappingSVintoQUBO} shows that an Ising spin Hamiltonian whose eigenstates correspond to lattice vectors in a certain region can be constructed in a straightforward manner. However this implicitly encodes the zero vector as the ground state of the Hamiltonian. Since the output of the SVP problem is expected to be non-zero, a technique to essentially find the Hamiltonian's first excited state must be utilised. Note that this is equivalent with finding a new Hamiltonian that has as ground state the first excited state of our initial Hamiltonian. In the case of VQE there exists a natural solution to the problem that can be accommodated fully at the classical loop (classical post-processing) of the optimization. This happens by defining a cost function that excludes the contribution of the zero vector eigenstate and consequently guides the optimiser towards the first excited state. However, such modification cannot be done to QAOA nor can be executed on a quantum annealer as the cost function is strictly defined by the optimization strategy itself.

The first approach one might consider is ignoring the $x\neq 0^n$ and proceed with trying to optimise for the ground state. 
There is a chance that the overlap of the first excited state with final ansatz state (aimed to match closely the ground state/zero-vector) is non-zero and that it gets sampled during the final ansatz measurements. This approach does not increase any requirements on quantum resources, however, especially as the number of qubits increases, it is not expected to yield satisfiable results. The difference in cost between the zero vector and the first excited state (the SVP) increases and the probability of obtaining the first excited state while targeting the ground state decays possibly exponentially fast.

Instead, a more accurate and in the long term, viable solution is to modify the Hamiltonian and impose a penalty for reaching the zero vector (ground state of the ``naive'' Hamiltonian). Introducing the extra constraint, and ensuring that the new Hamiltonian is still QUBO (so that we can run QAOA or quantum annealing) requires to introduce auxiliary variables and thus requires more qubits. In our case, we succeed to obtain a theoretical guarantee that the optimization result converges to the shortest lattice vector in the limit of optimization parameters for the cost of introducing extra $n-2$ qubits for a lattice with rank $n$. Such approach hence roughly reduces the abilities of QAOA algorithm or quantum annealer device by a half compared to the VQE algorithm but allows us to benefit from a wider range of quantum variational approaches and hence we include it here for the reference. A trivial solution would be to penalise certain assignment of QUBO binary variables that makes the QUBO formulation zero.  As we will argue below, given $m$ QUBO binary variables, the approach would require to introduce $m-2$ new binary variables. Note that $m\geq n$ is the number of binary variables in QUBO. If we let $n$ be rank of a lattice we can see that $m\geq n$ as shown in Section \ref{sec:mappingSVintoQUBO} where $m\approx n\log(2a)$ if each $x_i$ has the same variable range $2a$ assigned. Suppose that instead of the mapping presented in Equation \ref{eq:conversion_simple} we instead map each integer variable as
\begin{align}\label{eq:svpformulation4}
	|x_i|  \leq & a\implies \\
	x_i =&-a+\zeta_i a+\omega_i(a+1)+\sum_{j=0}^{\lfloor\log (a-1)\rfloor-1}2^j\tilde{x}_{ij}\nonumber\\
	& +(a-2^{\lfloor\log (a-1)\rfloor})\tilde{x}_{i,\lfloor\log (a-1)\rfloor}\nonumber
\end{align}
by introducing new binary variables $\zeta_i$, $\omega_i$ and $\{\tilde{x}_{ij}\}_{0\leq j \leq \lfloor\log a\rfloor}$ for eachi $x_i$.The encoding (\ref{eq:svpformulation4}) allows us to encode penalization of zero state with much less qubits. It is easy to observe that for the last two terms of (\ref{eq:svpformulation4}) it holds that
{\scriptsize\[
0\leq\sum_{j=0}^{\lfloor\log (a-1)\rfloor-1}2^j\tilde{x}_{ij}+(a-2^{\lfloor\log (a-1)\rfloor})\tilde{x}_{i,\lfloor\log (a-1)\rfloor} \leq a-1
\]}
and consequently
\[
x_i=0 \implies \zeta_i = 1
\]
Hence the penalization of the case where all integer variables are penalised for being zero $\forall x_i=0$ is equivalent to penalization of the case where the same number of binary variables are one $\forall \zeta_i=1$. The penalization term then becomes (\ref{eq:penalization1})
\begin{equation}\label{eq:penalization1}
	P\prod\zeta_i
\end{equation}
where $P$ is the penalty value. The term (\ref{eq:penalization1}) can be encoded as (\ref{eq:penalization2}) where $\{z_i\}_{1\leq i \leq n}$ are extra auxiliary binary variables with bijective correspondence to $\{\zeta_i\}_{0\leq i \leq n}$.
\begin{equation}\label{eq:penalization2}
	P\cdot \left(1+\sum_{i=1}^nz_i\cdot \left(-(1-\zeta_i)+\sum_{k=i+1}^n(1-\zeta_k)\right)\right).
\end{equation}
\begin{proof}[(\ref{eq:penalization2}) encodes constraint (\ref{eq:penalization1})]
	Let $x_i=1-\zeta_i$, $\tau_i=-x_i+\sum_{k=i+1}x_k$ and let $z_i=1\iff \tau_i<0$. Then it is easy to see that values of $z_1,\ldots,z_n$ minimise (\ref{eq:penalization2}). If $\forall x_i=0$, the whole expression is equal to $P$ as required. Otherwise, let $j_1,...,j_m$ be positions of $m\leq n$ binary ones in the bitstring $x_1 x_2 \ldots x_n$ sorted in the increasing order. Then $\tau_{j_m}=-1$ and $\tau_{j_l}\geq 0$ for $1\leq l<m$. Hence $\forall_{i\neq j_m} z_i=0$, $z_{j_m}=1$ and the whole expression is $0$, hence no penalty is imposed.
\end{proof}

Observe that setting $z_n=1$ and $z_{n-1}=\zeta_n$ does not change the global minimum of (\ref{eq:penalization2}). Hence $n-2$ additional binary variables $\{z_i\}_{1\leq i \leq n-2}$ are needed to encode the penalization term for an $n$-rank lattice. The quadratic unconstrained binary optimization problem formulation that can be trivially mapped to Ising spin Hamiltonian hence becomes
\begin{align}\label{eq:qubo}
  & \min_{\substack{x_1,\ldots,x_n\\
			z_1,...,z_{n-2}	\\
  }}\sum_{i=1}^n x_i\cdot G_{ii}\\
  & + 2\sum_{1\leq i<j\leq n} x_i\cdot x_j \cdot G_{ij}\nonumber\\
  &+ L\cdot \left(1+\sum_{i=1}^n z_i\cdot \left(-(1-\zeta_i)+\sum_{k=i+1}^n(1-\zeta_k)\right)\right)\nonumber
\end{align}
where ${x_i}$ is encoded as in \cref{eq:svpformulation4}. Number of binary variables needed to represent $\mathbf{x_i}$ bounded by $a_i$ is $\lfloor\log a_i\rfloor+3$. Hence the total number of binary variables in \cref{eq:qubo} is
\begin{align*}
  &n_{\text{bin\_vars\_constrained}}\\
  &=\sum_{i=1}^{n}(\lfloor\log a_i\rfloor+3)+(n-2)=4n-2+\sum_{i=1}^{n}\lfloor\log a_i\rfloor.
\end{align*}


%% file: appendix_naive_qubit_assignment.tex
As discussed in \cref{sec:compositeSVP,sec:exp_framework}, 
the bounds determined in \cref{sec:optBound} are not the best choice for small lattice instances. Given $n$ dimensional lattice $\mathcal{L}$ and $m$ available qubits, choosing a uniform distribution of qubits over elements of the coefficient vector $x$ turns out to achieve a high probability of encoding the shortest non-zero lattice vector as the ground state of a problem Hamiltonian when $m$ is sufficiently large. \Cref{fig:probGroundStates} depicts the probabilities of encoding the shortest non-zero lattice vector as a Hamiltonian's ground state given the number of available qubits and the lattice rank. The results were averaged over 1024 instances prepared using the same methodology as for the quantum emulation experiments  described in \cref{sec:exp_framework}. Three approaches have been compared:
\begin{description}
\item[blue] distributing $\floor{\frac{m}{n}}$ qubits to each of the coefficient of $x$,
\item[red] randomly distributing $\floor{\frac{m}{n}}$ qubits to $n\times\floor{\frac{m}{n}}$, coefficients of $x$ and distributing $\floor{\frac{m}{n}}+1$ to the remaining coefficients of $x$
\item[green] using \cref{lem:bound_xi_precise_dual} with $A=\gh(\cL)$ to obtain a discrete distribution of the bounds over coefficients of $x$; the distribution is then scaled such that at most $m$ qubits were distributed.
\end{description}



%% file: appendix_cvar.tex
We chose the value of $\alpha=0.175$ to use in the $CVaR_\alpha$ VQE algorithm in our quantum emulation experiments in Section \ref{sec:expResults}. Based on an experiment involving 1024 instances of rank 16 lattices prepared as in Section \ref{sec:exp_framework} this value gave the best results. The value of $\alpha$ was varied incrementally in small steps. Figures \ref{fig:sub1} and \ref{fig:sub2} depict the mean and the median overlaps of the final ansatz state with Hamiltonian's ground state respectively. We can observe that despite positive linear correlation between the mean overlap and parameter $\alpha$, the median overlap peaks sharply at $\alpha\approx0.175$, an indication of a rightly-skewed distribution. In other words, we have seen significantly less instances with high overlap as $\alpha>\approx 0.4$ tends towards one, although when such overlaps happen, they are high enough so that their mean increases with $\alpha$. The inverse of the overlap is the expected number of measurement samples of the final ansatz state to obtain the ground state. Hence a compromise must have been made to achieve as many as possible overlaps which were sufficiently high in the sense that the ground state could be obtained from them by a reasonable number of measurements. In order to make a fair comparison, we considered the probabilities of sampling the ground state if 5000 measurements of the final ansatz were to be performed as depicted in Figure \ref{fig:sub3}. We chose the value $\alpha=0.175$ as it achieved around $78\%$ probability of finding the Hamiltonian's ground state and at the same time it achieves the highest median of overlaps (around 0.006). Note that the non-$CVaR_\alpha$ VQE version, i.e. when $\alpha=1$ resulted in only $21\%$ probability of finding the Hamiltonian's ground state.

A final thing to note is that when one uses CVaR, the optimiser has incentive to find a quantum state that has $\alpha$ overlap with the true ground state, but no incentive to increase the overlap to higher values than $\alpha$, since anything above the $\alpha$-tail is irrelevant for computing the cost. Therefore, while it has been demonstrated that small values of $\alpha$ may increase the chances of finding the solution after a fixed number of shots/measurements of the final state is carried out, the actual (mean) overlap does not necessarily increase (at least not to values above $\alpha$). This was one of the motivation to look modify CVaR and define an Ascending-CVaR \cite{ascending_cvar}.

%% file: appendix_proofs.tex
\section{Proof of \Cref{cor:ortho_defect_hkz}}
\label{sec:proof_ortho_defect_hkz}

	Let $\delta$ be the orthogonality defect. Then
	\begin{align*}
		\log_2\delta
			&\leqslant \tfrac{n}{2}\log_2\gamma_n
				+\tfrac{1}{2}\log_2\tfrac{(n+3)!}{3!}-n\\
			&\leqslant \tfrac{n}{2}\log_2(\tfrac{1}{8}n+\tfrac{6}{5})-n
				+\tfrac{1}{2}\log_2(n+3)!+O(1)\\
			&=\tfrac{n}{2}\log_2(n)-\tfrac{5}{2}n
				+\tfrac{1}{2}\log_2(n+3)!+O(1)\\
			&=\tfrac{n}{2}\log_2(n)-\tfrac{5}{2}n
				+\tfrac{1}{2}\log_2n!+O(\log n)\\
			&\leqslant\tfrac{n}{2}\log_2(n)-\tfrac{5}{2}n
				+\tfrac{1}{2}\log_2\left(\sqrt{2\pi n}\left(\tfrac{n}{e}\right)^ne^{\tfrac{1}{12n}}\right)+O(\log n)\\
			&=\tfrac{n}{2}\log_2(n)-\tfrac{5}{2}n
				+\tfrac{n}{2}\log_2n
				-\tfrac{n}{2}\log_2e+O(\log n)
	\end{align*}
where we used \cite{Robbins55} at the second to last line.

%% file: experimental_results/probGroundState.tex
\begin{figure*}[tbp]
	\centering
	\begin{tikzpicture}
		\begin{groupplot}[
			group style={
				group name=my plots,
				group size=4 by 3,
				ylabels at=edge left
			},
			footnotesize,
			width=4.5cm,
			height=4cm,
			tickpos=left,
			ytick align=outside,
			xtick align=outside,
			enlarge x limits=false,
			ymin=-0.1,ymax=1.1
			]
			\definecolor{mygr}{HTML}{2DA10B}
			
			\nextgroupplot[title={Available qubits: 40}, ylabel=Probability]
			
			\addplot[blue, dashed] table{experimental_results/quantum_exp_data/uniform_pgfplot40.txt};
			\addplot[red, dashed] table{experimental_results/quantum_exp_data/uniform_random_pgfplot40.txt};
			\addplot[mygr, dashed] table{experimental_results/quantum_exp_data/dualbounded_pgfplot40.txt};
			
			\nextgroupplot[title={Available qubits: 60}]
			\addplot[blue, dashed] table{experimental_results/quantum_exp_data/uniform_pgfplot60.txt};
			\addplot[red, dashed] table{experimental_results/quantum_exp_data/uniform_random_pgfplot60.txt};
			\addplot[mygr, dashed] table{experimental_results/quantum_exp_data/dualbounded_pgfplot60.txt};
			
			\nextgroupplot[title={Available qubits: 80}]
			\addplot[blue, dashed] table{experimental_results/quantum_exp_data/uniform_pgfplot80.txt};
			\addplot[red, dashed] table{experimental_results/quantum_exp_data/uniform_random_pgfplot80.txt};
			\addplot[mygr, dashed] table{experimental_results/quantum_exp_data/dualbounded_pgfplot80.txt};
			
			\nextgroupplot[title={Available qubits: 100}]
			\addplot[blue, dashed] table{experimental_results/quantum_exp_data/uniform_pgfplot100.txt};
			\addplot[red, dashed] table{experimental_results/quantum_exp_data/uniform_random_pgfplot100.txt};
			\addplot[mygr, dashed] table{experimental_results/quantum_exp_data/dualbounded_pgfplot100.txt};
			
			\nextgroupplot[title={Available qubits: 120}, ylabel=Probability]
			\addplot[blue, dashed] table{experimental_results/quantum_exp_data/uniform_pgfplot120.txt};
			\addplot[red, dashed] table{experimental_results/quantum_exp_data/uniform_random_pgfplot120.txt};
			\addplot[mygr, dashed] table{experimental_results/quantum_exp_data/dualbounded_pgfplot120.txt};
			
			\nextgroupplot[title={Available qubits: 140}]
			\addplot[blue, dashed] table{experimental_results/quantum_exp_data/uniform_pgfplot140.txt};
			\addplot[red, dashed] table{experimental_results/quantum_exp_data/uniform_random_pgfplot140.txt};
			\addplot[mygr, dashed] table{experimental_results/quantum_exp_data/dualbounded_pgfplot140.txt};
			
			\nextgroupplot[title={Available qubits: 160}]
			\addplot[blue, dashed] table{experimental_results/quantum_exp_data/uniform_pgfplot160.txt};
			\addplot[red, dashed] table{experimental_results/quantum_exp_data/uniform_random_pgfplot160.txt};
			\addplot[mygr, dashed] table{experimental_results/quantum_exp_data/dualbounded_pgfplot160.txt};
			
			\nextgroupplot[title={Available qubits: 180}]
			\addplot[blue, dashed] table{experimental_results/quantum_exp_data/uniform_pgfplot180.txt};
			\addplot[red, dashed] table{experimental_results/quantum_exp_data/uniform_random_pgfplot180.txt};
			\addplot[mygr, dashed] table{experimental_results/quantum_exp_data/dualbounded_pgfplot180.txt};
			
			\nextgroupplot[title={Available qubits: 200}, xlabel=Lattice rank, ylabel=Probability]
			\addplot[blue, dashed] table{experimental_results/quantum_exp_data/uniform_pgfplot200.txt};
			\addplot[red, dashed] table{experimental_results/quantum_exp_data/uniform_random_pgfplot200.txt};
			\addplot[mygr, dashed] table{experimental_results/quantum_exp_data/dualbounded_pgfplot200.txt};
			
			\nextgroupplot[title={Available qubits: 220}, xlabel=Lattice rank]
			\addplot[blue, dashed] table{experimental_results/quantum_exp_data/uniform_pgfplot220.txt};
			\addplot[red, dashed] table{experimental_results/quantum_exp_data/uniform_random_pgfplot220.txt};
			\addplot[mygr, dashed] table{experimental_results/quantum_exp_data/dualbounded_pgfplot220.txt};
			
			\nextgroupplot[title={Available qubits: 240}, xlabel=Lattice rank]
			\addplot[blue, dashed] table{experimental_results/quantum_exp_data/uniform_pgfplot240.txt};
			\addplot[red, dashed] table{experimental_results/quantum_exp_data/uniform_random_pgfplot240.txt};
			\addplot[mygr, dashed] table{experimental_results/quantum_exp_data/dualbounded_pgfplot240.txt};
			
			\nextgroupplot[title={Available qubits: 260}, xlabel=Lattice rank]
			\addplot[blue, dashed] table{experimental_results/quantum_exp_data/uniform_pgfplot260.txt};
			\addplot[red, dashed] table{experimental_results/quantum_exp_data/uniform_random_pgfplot260.txt};
			\addplot[mygr, dashed] table{experimental_results/quantum_exp_data/dualbounded_pgfplot260.txt};

		\end{groupplot}
	\end{tikzpicture}
	\caption{Probability of encoding the shortest lattice vector as an eigenvector of problem Hamiltonian using \textbf{blue:} uniform distribution of qubits, \textbf{red:} uniform distribution of qubits and random distribution of the remaining qubits, \textbf{green:} scaled distribution of qubits according to Lemma \ref{lem:bound_xi_precise_dual} with A being a gaussian heuristics corresponding to the lattice instance}\label{fig:probGroundStates}
\end{figure*}
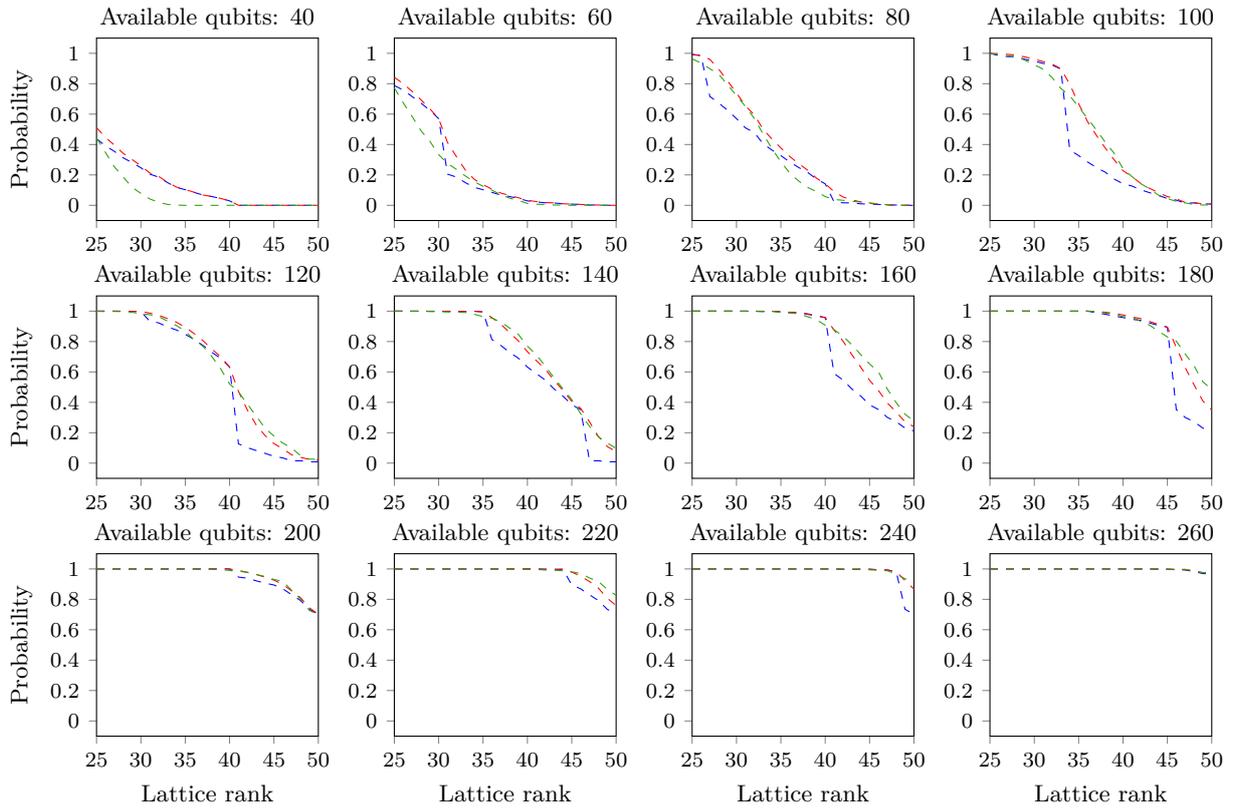

%% file: appendix_cvar_fig.tex
\begin{figure*}[tbp]
	\caption{Experimental determination of CVaR's parameter $\alpha$}\label{fig:cvar}
	
	\begin{subfigure}{.5\linewidth}
		\centering
		\begin{tikzpicture}			
			\begin{axis}[ymin=0, ymax=0.61, xmin=0, xmax=1.01, xlabel=Parameter $\alpha$, ylabel=Overlap,height=5cm,
				yticklabel style={
					/pgf/number format/fixed,
					/pgf/number format/precision=4,
				},]
				\addplot[blue, solid, mark=o]
				plot [error bars/.cd, y dir = both, y explicit]
				table[x =x, y =y, y error =e]{experimental_results/quantum_exp_data/cvar_hit_rates.txt};	
			\end{axis}	
		\end{tikzpicture}
		\caption{Averaged overlap with the shortest lattice vector as a function of parameter $\alpha$ with standard deviations.}	\label{fig:sub1}
	\end{subfigure}
	\begin{subfigure}{.5\linewidth}
		\centering
		\begin{tikzpicture}			
			\begin{axis}[ymin=0, ymax=0.007, xmin=0, xmax=1.01, xlabel=Parameter $\alpha$, ylabel=Overlap,height=5cm,
				scaled y ticks=false,
				yticklabel style={
					/pgf/number format/fixed,
					/pgf/number format/precision=4,
				}]
				\addplot[blue, solid, mark=o] table{experimental_results/quantum_exp_data/cvar_medians.txt};
			\end{axis}	
		\end{tikzpicture}
		\caption{Median overlap with the shortest lattice vector as a function of parameter $\alpha$.}\label{fig:sub2}
	\end{subfigure}\\[1ex]
	\begin{subfigure}{\linewidth}
		\centering
		\begin{tikzpicture}			
			\begin{axis}[xlabel=Parameter $\alpha$, ylabel=Prob.~finding ground state,ymin=0.1,ymax=0.85,xmin=0,xmax=1.01,ytick align=outside,height=5cm,
				xtick align=outside,
				scaled y ticks=false,
				yticklabel style={
					/pgf/number format/fixed,
					/pgf/number format/precision=3,
				}]
				\addplot[blue, solid, mark=o] table{experimental_results/quantum_exp_data/cvar_5000samples.txt};
			\end{axis}	
		\end{tikzpicture}
		\caption{Probability of sampling the ground state in 5000 samples using CVaR version of VQE as a function of parameter $\alpha$. Note that $\alpha=1$ corresponds to the standard non-CVaR version of VQE.
		}\label{fig:sub3}
	\end{subfigure}
\end{figure*}
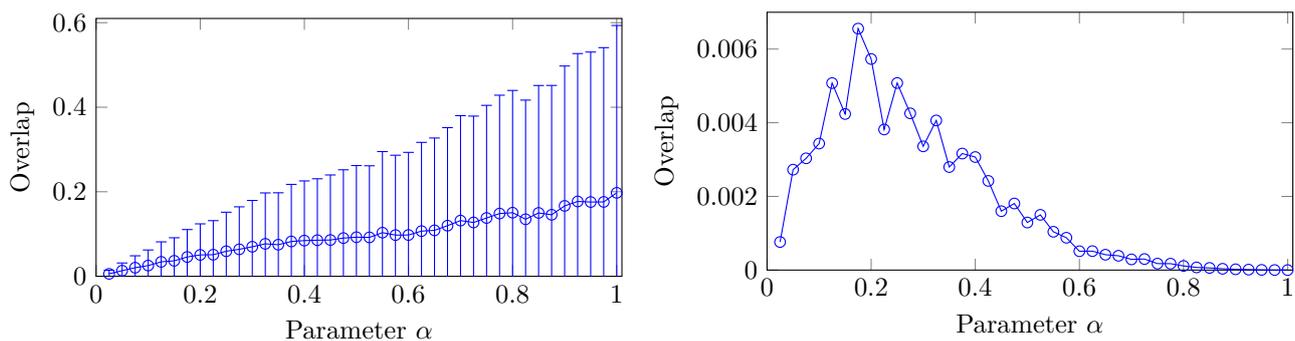